\documentclass[12pt,round]{article}

\usepackage[a4paper,margin=2.3cm]{geometry}
\usepackage{graphicx}
\usepackage{subcaption}
\usepackage{amsmath}
\usepackage{amssymb}
\usepackage{bm}
\usepackage{natbib}
\usepackage{booktabs}
\usepackage{listings}
\usepackage{url}
\usepackage{placeins}
\usepackage{authblk}
\usepackage{caption}
\usepackage{microtype}

\makeatletter
\renewcommand\section{\@startsection{section}{1}{\z@}%
  {-2.0ex plus -0.5ex minus -.2ex}%
  {1.0ex plus .2ex}%
  {\normalfont\large\bfseries}}
\renewcommand\subsection{\@startsection{subsection}{2}{\z@}%
  {-1.5ex plus -0.4ex minus -.2ex}%
  {0.7ex plus .2ex}%
  {\normalfont\normalsize\bfseries}}
\makeatother

\newtheorem{proposition}{Proposition}

\title{\Large\bfseries Covariance shrinkage for cosmological inference with Sellentin--Heavens-type likelihoods}

\author[1]{Raffaele Mattera}
\affil[1]{\small Department of Mathematics and Physics, University of Campania ``Luigi Vanvitelli'', Caserta, Italy\\
\texttt{raffaele.mattera@unicampania.it}}

\date{}

\begin{document}

\maketitle

\vspace{-1.5em}

\begin{abstract}
\noindent
Covariance matrices used in astronomical and cosmological parameter inference are often estimated from a finite number of simulations. The resulting covariance uncertainty affects inferential results. We study covariance regularisation from the perspective of likelihood-based inference with simulation-estimated covariance matrices. We first analyse scalar covariance scaling under the Gaussian plug-in likelihood and the covariance-marginalised Sellentin--Heavens likelihood. We then introduce a shrinkage-intensity formulation in which the amount of covariance regularisation is treated as an auxiliary inferential quantity rather than as a fixed plug-in tuning parameter. We define an expected negative log-likelihood loss and derive likelihood-optimal scalar covariance scalings. We then consider a rotation-equivariant linear shrinkage family in which the sample covariance eigenvalues are shrunk towards their grand mean. A prior distribution is assigned to the shrinkage intensity, the likelihood induces its posterior distribution, and the final parameter posterior is obtained by marginalising over the shrinkage intensity. For scalar covariance corrections, Hartlap covariance-side scaling is recovered as the likelihood-loss optimum under the Gaussian plug-in likelihood. In contrast, under the Sellentin--Heavens likelihood, the optimal scalar correction is the unscaled sample covariance. Simulation experiments confirm that shrinkage substantially improves covariance conditioning, while marginalisation over the shrinkage intensity propagates uncertainty about the amount of regularisation into posterior inference. Scalar covariance corrections are likelihood-dependent. Under the Sellentin--Heavens likelihood, additional Hartlap-type global scaling is not favoured, but structural covariance regularisation can still improve inference when the sample covariance is noisy. Treating the shrinkage intensity as an auxiliary inferential parameter provides a simple way to combine covariance-marginalised likelihood inference with regularisation of simulation-estimated covariance matrices.
\end{abstract}

\vspace{0.75em}

\vspace{1.5em}

\section{Introduction}
\label{sec:introduction}

In cosmological parameter inference, the objective is to estimate a vector of model
parameters \(\theta\), together with its associated uncertainty, from an observed
data vector. Let \(x_o\in\mathbb{R}^{p}\) denote the observed summary-statistic data
vector and let \(\mu_\theta\in\mathbb{R}^{p}\) be the model prediction associated with
\(\theta\). Under the standard Gaussian approximation, the sampling distribution of
the observed data vector is written as $x_o\mid\theta,\Sigma
\sim
\mathcal{N}_p(\mu_\theta,\Sigma)$, where \(\Sigma\) is the covariance matrix of the data vector. The corresponding
likelihood, viewed as a function of \(\theta\) for the fixed observed vector \(x_o\),
is
\begin{equation}
\mathcal{L}_G(\theta)
=
(2\pi)^{-p/2}|\Sigma|^{-1/2}
\exp\left\{
-\frac{1}{2}
(x_o-\mu_\theta)^\top
\Sigma^{-1}
(x_o-\mu_\theta)
\right\}.
\label{eq:gaussian_likelihood}
\end{equation}
Given a prior density \(\pi(\theta)\), Bayesian inference is based on the posterior
density
\begin{equation}
\pi(\theta\mid x_o,\Sigma)
=
\frac{
\mathcal{L}_G(\theta)\pi(\theta)
}{
\int_{\Theta}
\mathcal{L}_G(\vartheta)\pi(\vartheta)\,d\vartheta
}.
\label{eq:posterior_gaussian}
\end{equation}
Point estimates may then be obtained from posterior summaries, such as the
posterior mean \(E(\theta\mid x_o,\Sigma)\), or from the maximum a posteriori
estimator. Uncertainty is described by the posterior distribution itself, for
example through marginal credible intervals or joint credible regions. Thus, both
parameter estimates and their uncertainty depend on the covariance matrix
\(\Sigma\), because \(\Sigma\) determines the metric by which discrepancies between
the observed data vector \(x_o\) and the model prediction \(\mu_\theta\) are weighted.

In an ideal analysis, both the mean vector \(\mu_\theta\) and the covariance matrix
\(\Sigma\) would be available analytically. In many astronomical and cosmological
applications, however, the covariance matrix is
estimated from a finite number of mock data vectors. If
\(x_1,\ldots,x_{n_s}\) denote such mock data vectors, the usual simulation-based
covariance estimator $S$ is
used. Under Gaussian simulations, \((n_s-1)S\sim W_p(\Sigma,n_s-1)\), where \(W_p\)
denotes the Wishart distribution. The covariance matrix entering the likelihood is
therefore itself a random estimate. The statistical impact of simulation-estimated covariance matrices is well
recognised in cosmological inference \citep[e.g. see][]{hartlap2007your,taylor2013putting}, as this directly affects both estimates and their uncertainty quantification.

A first problem is that, given mock simulations, although the sample covariance estimator
\(S\) is unbiased for the true covariance \(\Sigma\) its inverse \(S^{-1}\) is a biased estimator
for the true precision matrix \(\Sigma^{-1}\). This matters because the precision
matrix enters the Gaussian likelihood through the quadratic form in
\eqref{eq:gaussian_likelihood}. A standard response in astrostatistics is the Hartlap correction \citep{hartlap2007your}, which rescales the inverse sample
covariance matrix in order to remove this bias under the Wishart sampling model. This correction addresses the bias of the precision estimator
while retaining the Gaussian likelihood \eqref{eq:gaussian_likelihood} and improves the inference.

However, debiasing the precision matrix does not entirely remove the
inferential consequences of covariance noise. \citet{dodelson2013effect}
showed that uncertainty in a simulation-estimated covariance matrix propagates
to parameter inference by inflating parameter variances, with a leading effect of order
\(1+p/n_s\), where \(p\) is the dimension of the data vector and \(n_s\) is
the number of simulations used to estimate the covariance matrix. This result
highlights a first difficulty, that is, covariance estimation becomes increasingly
important when the data-vector dimension is not small relative to the number of
available simulations. In such regimes, the sample covariance may be poorly
conditioned, its inverse may be unstable, and Gaussian plug-in inference may
become unreliable.  To address this high-dimensional covariance-estimation problem, several
strategies have been proposed in the cosmological literature. These include
smooth analytical or semi-analytical covariance models calibrated with fewer
mocks \citep{o2016large,o2019large}, shrinkage estimators
\citep{joachimi2017non}, precision-matrix expansions around analytical
covariance models \citep{friedrich2018precision}, surrogate-assisted covariance
estimation \citep{chartier2021carpool,chartier2022carpool}, and
compression-based methods that reduce the dimensionality of the data vector and
therefore the impact of covariance-estimation noise
\citep{philcox2021fewer,sugiyama2025data}. These methods demonstrate that covariance estimation is not a secondary
technical detail in cosmological inference: the way in which the covariance
matrix is estimated, regularised or compressed can directly affect posterior
geometry and parameter constraints. The present work follows this line of
reasoning, but shifts the focus from constructing a single improved covariance
estimate to incorporating covariance regularisation into the likelihood used for
parameter inference.

Indeed, \citet{dodelson2013effect} also highlight that covariance noise is not only a
matrix-estimation problem. The estimated covariance matrix enters the likelihood
through the precision matrix and the geometry of the parameter posterior.
Therefore, an alternative strategy is to modify the likelihood itself. Under
Gaussian mock data vectors and the independence Jeffreys prior for \(\Sigma\),
marginalising over the unknown true covariance matrix conditional on the
simulation estimate leads to the \citet{sellentin2015parameter} covariance-marginalised
likelihood. This likelihood replaces the Gaussian
plug-in likelihood by a heavier-tailed likelihood that accounts explicitly for
the finite number of simulations used to estimate the covariance matrix. This likelihood-level viewpoint was further developed by
\citet{percival2022matching}, who showed that different priors on the unknown
covariance matrix lead to different \(t\)-type posteriors.
This shows that the covariance correction depends on the inferential
target, whether this target is precision-matrix debiasing, covariance
marginalisation or a coverage calibration.

The present paper addresses a complementary question. Rather than changing the
prior on the unrestricted covariance matrix in order to modify the resulting
covariance-marginalised posterior, we regularise the simulation-estimated
covariance matrix used inside the likelihood. Specifically, we combine the Sellentin--Heavens covariance-marginalised
likelihood with a shrinkage family,
\[
\widehat{\Sigma}_\alpha
=
\alpha S+(1-\alpha)\tau I_p,
\qquad
\tau=p^{-1}\operatorname{tr}(S),
\]
and treat the shrinkage intensity \(\alpha\) as an auxiliary inferential
quantity. A prior is assigned to \(\alpha\), the likelihood induces its posterior
distribution, and the final posterior of the scientific parameters is obtained by
marginalising over \(\alpha\). Therefore, the aim is not to construct a new covariance
shrinkage estimator under a standalone matrix loss \citep[as in, for instance,][]{ledoit2004well,fan2013large,mattera2025improved}, but to understand how covariance
shrinkage can be incorporated into likelihood-based parameter inference. 

Therefore, the contribution is twofold. First, we combine covariance-marginalised likelihood inference with high-dimensional covariance regularisation by using a shrinkage covariance matrix inside the Sellentin--Heavens likelihood. In this way, finite-simulation covariance uncertainty is handled through the heavier-tailed likelihood, while the instability of the simulation-estimated covariance matrix is addressed through shrinkage. Second, the shrinkage intensity is not fixed by an external matrix loss or inserted as a plug-in estimate. Instead, \(\alpha\) is treated as an auxiliary inferential quantity: it is assigned a prior, learned through the likelihood, and marginalised in the final posterior of the scientific parameters. This propagates uncertainty about the amount of regularisation directly into parameter inference.

To further motivate this construction, we first consider the simplest possible
regularisation class, namely global scalar rescaling of the sample covariance,
\[
\widehat{\Sigma}_c=cS,\qquad c>0.
\]
This class changes only the overall scale of the covariance matrix. It inflates
or deflates all eigenvalues by the same factor and therefore leaves the relative
eigenvalue structure of \(S\) unchanged. Scalar rescaling is consequently useful
for studying global likelihood calibration, but it cannot correct noisy
eigenvalue dispersion or poor conditioning.

This scalar analysis provides the bridge to shrinkage. We show that, under the Gaussian
plug-in likelihood, the likelihood-loss optimal scalar correction recovers the
Hartlap covariance-side scaling. We then demonstrate that, under the Sellentin--Heavens likelihood, the
optimal scalar correction is instead \(c=1\), meaning that the unscaled sample
covariance is already optimal within the scalar class \(\{cS:c>0\}\). This does
not imply that the sample covariance is optimal among all regularised covariance
estimators. It only shows that, once covariance uncertainty has been incorporated
through the Sellentin--Heavens likelihood, further improvement should not be
sought through an additional global multiplicative factor. The remaining problem
is structural: the sample covariance may still have noisy eigenvalues and poor
conditioning. This motivates the move from scalar rescaling to shrinkage, which
regularises the covariance structure by pulling the eigenvalues of \(S\) towards
a stable target. 

We validate the proposed approach through
Monte Carlo simulations. The accompanying R and Python implementations are
available at \url{https://github.com/raffmattera/cosmo-shrinkage-inference}.

The remainder of the paper is organised as follows.
Section~\ref{sec:pre} fixes the notation and main
concepts used throughout the paper. Section~\ref{sec:theory_scaling} derives
likelihood-loss optimal scalar covariance corrections under the Gaussian and
Sellentin--Heavens likelihoods. Section~\ref{sec:shrinkage} introduces the Bayesian
shrinkage-intensity formulation and its posterior computation. Section~\ref{sec:simulation}
presents simulation studies on scalar likelihood loss, posterior shrinkage behaviour
and posterior calibration. Section~\ref{sec:conclusion} concludes.

\section{Preliminaries}
\label{sec:pre}

This section fixes the notation for covariance-estimated likelihoods. The aim is to clarify the statistical
objects used in the theoretical and simulation results below.

\subsection{Observed data, mock data vectors, and covariance estimation}
\label{subsec:mock_covariance}

Let \(x_o\in\mathbb{R}^{p}\) denote the observed summary-statistic data vector and
let \(\mu_\theta\in\mathbb{R}^{p}\) denote the model prediction associated with the
parameter vector \(\theta\). Conditional on \(\theta\) and on the covariance matrix
\(\Sigma\), the Gaussian data-vector model is
\begin{equation}
x_o\mid\theta,\Sigma
\sim
\mathcal{N}_p(\mu_\theta,\Sigma).
\label{eq:data_vector_model}
\end{equation}
The covariance matrix \(\Sigma\) is estimated from \(n_s\) independent mock data
vectors \(x_1,\ldots,x_{n_s}\), each having the same dimension and interpretation as
\(x_o\). The mock data vectors are synthetic realisations of the experiment and are
used only to estimate the covariance structure of the data vector. The sample covariance estimator is
\begin{equation}
S
=
\frac{1}{n_s-1}
\sum_{i=1}^{n_s}
(x_i-\bar{x})(x_i-\bar{x})^\top,
\qquad
\bar{x}
=
\frac{1}{n_s}
\sum_{i=1}^{n_s}x_i .
\label{eq:sample_covariance}
\end{equation}
Under Gaussian simulations,
\begin{equation}
(n_s-1)S
\sim
W_p(\Sigma,n_s-1).
\label{eq:wishart_sampling}
\end{equation}
Throughout the paper, \(n_s\) denotes the number of mock data vectors used to
estimate the covariance matrix, while \(p\) denotes the dimension of the observed
data vector.

\subsection{Gaussian plug-in likelihood and Hartlap correction}
\label{subsec:hartlap_scaling}

The Gaussian plug-in likelihood replaces the unknown covariance matrix \(\Sigma\)
by the simulation-based estimate \(S\) in \eqref{eq:gaussian_likelihood}.
Although \(S\) is unbiased for \(\Sigma\), the inverse sample covariance is biased.
For \(n_s>p+2\),
\begin{equation}
E(S^{-1})
=
\frac{n_s-1}{n_s-p-2}
\Sigma^{-1}.
\label{eq:inverse_covariance_bias}
\end{equation}
The Hartlap correction therefore replaces \(S^{-1}\) by
\begin{equation}
\widehat{\Omega}_{H}
=
\frac{n_s-p-2}{n_s-1}S^{-1}.
\label{eq:hartlap_precision}
\end{equation}
This correction debiases the inverse covariance estimator under the Wishart sampling
model, but the likelihood remains Gaussian in form.

\subsection{Sellentin--Heavens covariance-marginalised likelihood}
\label{subsec:sh_likelihood}

The Sellentin--Heavens approach treats the true covariance matrix \(\Sigma\) as
unknown and integrates it out conditional on the simulation-based estimate \(S\).
The covariance-marginalised likelihood is
\begin{equation}
p(x_o\mid\mu_\theta,S,n_s)
=
\int
p(x_o\mid\mu_\theta,\Sigma)
p(\Sigma\mid S,n_s)
\,d\Sigma .
\label{eq:covariance_marginalisation}
\end{equation}
Under Gaussian mock data vectors and the independence Jeffreys prior for
\(\Sigma\), this integral yields
\begin{equation}
\mathcal{L}_{SH}(\theta)
=
\bar{c}_p
|S|^{-1/2}
\left[
1+
\frac{
(x_o-\mu_\theta)^\top S^{-1}(x_o-\mu_\theta)
}
{n_s-1}
\right]^{-n_s/2},
\label{eq:sh_likelihood}
\end{equation}
where \(\bar{c}_p\) is the normalising constant and \(n_s>p\). The heavier tails of
\eqref{eq:sh_likelihood} encode the uncertainty associated with estimating the
covariance matrix from finitely many simulations.
\section{Optimal covariance scaling}
\label{sec:theory_scaling}

Let \(L_c(x_o\mid \theta,S)\) be a working likelihood constructed using a scaled covariance estimator \(\widehat{\Sigma}_c=cS\), with \(c>0\). We define the expected negative log-likelihood loss as
\begin{equation}
\mathcal{R}(c;L)
=
-
E_{x_o,S}
\left[
\log L_c(x_o\mid \theta_0,S)
\right],
\label{eq:generic_loss_c}
\end{equation}
where \(\theta_0\) denotes the true parameter value and the expectation is taken with respect to both the observed data vector and the simulated covariance estimate. The likelihood-loss optimal covariance scaling is
\[
c_L^\star
=
\arg\min_{c>0}
\mathcal{R}(c;L).
\]
This criterion makes explicit that the optimal covariance correction depends on the likelihood family.

\subsection{Gaussian likelihood}

Consider first the Gaussian plug-in likelihood \eqref{eq:gaussian_likelihood} with covariance \(cS\). Up to constants independent of \(c\), its expected negative log-likelihood loss is
\[
\mathcal{R}_G(c)
=
\frac{1}{2}
E_{x_o,S}
\left[
\log|cS|
+
(x_o-\mu_{\theta})^\top(cS)^{-1}(x_o-\mu_{\theta})
\right].
\]

\begin{proposition}[Gaussian likelihood-loss optimal scaling]
Assume that \(x_o\sim\mathcal{N}_p(\mu_{\theta},\Sigma)\) and \((n_s-1)S\sim W_p(\Sigma,n_s-1)\), with \(n_s>p+2\). Then \(\mathcal{R}_G(c)\) is uniquely minimised at
\begin{equation}
c_G^\star
=
\frac{n_s-1}{n_s-p-2}.
\label{eq:c_gaussian}
\end{equation}
Equivalently,
\[
(c_G^\star S)^{-1}
=
\frac{n_s-p-2}{n_s-1}S^{-1}.
\]
Thus, the Hartlap-corrected precision matrix is obtained as the likelihood-loss optimal precision matrix under a Gaussian working likelihood.
\end{proposition}

\noindent The proof is shown in Appendix \ref{app:proof_prop1}, using the technical results in Appendix \ref{app:multivariate_results}. Proposition~1 provides a decision-theoretic interpretation of the Hartlap correction. Hartlap scaling is optimal when the working likelihood is Gaussian and the loss is the expected negative Gaussian log-likelihood. However, this does not imply that the same scaling remains optimal once covariance uncertainty is incorporated directly into the likelihood.

\subsection{Sellentin--Heavens likelihood}
\label{subsec:sh_scaling}

We now consider the Sellentin--Heavens functional form evaluated with the scaled
covariance \(cS\). For \(c\neq 1\), this should be understood as a scalar
working family used to study global covariance calibration, not as a new exact
Wishart--Jeffreys marginalisation. Up to constants independent of \(c\), the
corresponding expected negative log-likelihood loss is
\begin{align}
& \mathcal{R}_{SH}(c)
=\\
&=
E_{x_o,S}
\left[
\frac{1}{2}\log|cS|
+
\frac{n_s}{2}
\log
\left\{
1+
\frac{
(x_o-\mu_{\theta})^\top(cS)^{-1}(x_o-\mu_{\theta})
}
{n_s-1}
\right\}
\right]. \nonumber  
\label{eq:risk_sh_scaled}
\end{align}

\begin{proposition}[Sellentin--Heavens likelihood-loss optimal scaling]
\label{prop:sh_scaling}
Assume that \(x_o\sim\mathcal{N}_p(\mu_{\theta},\Sigma)\) and
\((n_s-1)S\sim W_p(\Sigma,n_s-1)\), with \(x_o\) and \(S\) independent and
\(n_s>p\). Consider the scalar covariance class \(\{cS:c>0\}\) inside the
Sellentin--Heavens working likelihood. Then the expected negative log-likelihood
loss \(\mathcal{R}_{SH}(c)\) is uniquely minimised at
\begin{equation}
c_{SH}^\star=1.
\label{eq:c_sh}
\end{equation}
Thus, under the Sellentin--Heavens likelihood, no additional global scalar
inflation of the sample covariance is optimal within the class \(\{cS:c>0\}\).
\end{proposition}

\noindent
The proof is shown in Appendix \ref{app:proof_prop2}, using the technical results in Appendix \ref{app:multivariate_results}. Proposition 2 shows that, under the Sellentin--Heavens
likelihood, the optimal scalar correction is no correction: the unscaled sample
covariance \(S\) is already calibrated within the class \(\{cS:c>0\}\). The
result \(c_{SH}^\star=1\) is specific to the Sellentin--Heavens likelihood
associated with the independence Jeffreys prior. Other covariance-marginalised
likelihoods, such as the frequentist-matching construction of
\citet{percival2022matching}, modify the power-law exponent of the resulting
\(t\)-type posterior and may therefore imply different scalar calibrations.

For the Sellentin--Heavens likelihood, the result \(c_{SH}^\star=1\) should be
read as a likelihood-level calibration statement. The novelty is not the
algebraic proof itself, but the interpretation: after covariance uncertainty has
been incorporated through the Sellentin--Heavens marginalisation, an additional
global Hartlap-type rescaling of \(S\) is not favoured within the scalar class
\({cS:c>0}\). The unscaled sample covariance is therefore the optimal
global-scale representative of this class under the assumptions leading to the
Sellentin--Heavens likelihood.

This conclusion does not claim that \(S\) is optimal among structured covariance
estimators. It only rules out a global multiplicative correction as the relevant
direction for further improvement. If improvement is still needed, it must act on
the internal covariance structure, for example through shrinkage, rather than on
the overall scale of the matrix.

\section{Covariance shrinkage}
\label{sec:shrinkage}

The previous section considered scalar covariance corrections of the form \(cS\).
Such corrections modify the global scale of the covariance matrix but leave its
relative eigenvalue structure unchanged. This is a restrictive class of
regularisation. When the dimension \(p\) of the data vector is not negligible
relative to the number of simulations \(n_s\), the sample covariance matrix can be
poorly conditioned, its eigenvalues can be overly dispersed, and its inverse can be
unstable. In this regime, the main difficulty is not only the global scale of the
covariance estimator, but also the noisy structure of its spectrum.

This motivates covariance shrinkage. In high-dimensional statistics, linear
shrinkage estimators combine the sample covariance matrix with a structured
low-variance target, often a scalar multiple of the identity matrix. Such estimators
can be interpreted as bias--variance trade-offs or, equivalently, as shrinking the
sample eigenvalues towards a common value. Classical \cite{ledoit2004well}-type estimators
select the shrinkage intensity by minimising a matrix loss, typically a Frobenius
risk. Stein-type \citep{dey1985estimation,fisher2011improved} and Haff-type \citep{haff1977minimax,haff1980empirical} estimators arise instead from invariant loss
functions, empirical Bayes arguments, or dominance results for covariance or
precision matrix estimation. These approaches are decision-theoretic or
likelihood-related in different senses, but their primary object is the estimation of
\(\Sigma\) or \(\Sigma^{-1}\).

The objective considered here is different. We use covariance shrinkage inside the
likelihood used for parameter inference and treat the amount of shrinkage as an auxiliary inferential quantity. A prior distribution is assigned to the shrinkage intensity, the likelihood induces its posterior distribution, and the shrinkage intensity is marginalised out. This allows
uncertainty in the amount of covariance regularisation to propagate into the final
posterior distribution of the model parameters.

\subsection{Shrinkage covariance family}
\label{subsec:shrinkage_family}

We consider the rotation-equivariant linear shrinkage family
\begin{equation}
\widehat{\Sigma}_\alpha
=
\alpha S
+
(1-\alpha)\tau I_p,
\qquad
\tau=\frac{1}{p}\operatorname{tr}(S),
\qquad
0<\alpha<1.
\label{eq:shrinkage_estimator}
\end{equation}
The case \(\alpha\approx1\) corresponds to weak shrinkage and gives a covariance
estimator close to the sample covariance \(S\), whereas smaller values of
\(\alpha\) shrink the covariance matrix towards the spherical target \(\tau I_p\). The meaning of shrinkage in \eqref{eq:shrinkage_estimator} is most transparent
from the spectral decomposition of \(S\). If
\[
S=Q\Lambda Q^\top,
\qquad
\Lambda=\operatorname{diag}(\lambda_1,\ldots,\lambda_p),
\]
then
\[
\widehat{\Sigma}_\alpha
=
Q
\operatorname{diag}
\left\{
\alpha\lambda_i+(1-\alpha)\bar{\lambda}
\right\}_{i=1}^{p}
Q^\top,
\qquad
\bar{\lambda}
=
\frac{1}{p}\sum_{i=1}^{p}\lambda_i .
\]
Thus, shrinkage leaves the sample eigenvectors unchanged and replaces each
sample eigenvalue \(\lambda_i\) by a convex combination of \(\lambda_i\) and the
average eigenvalue \(\bar{\lambda}\). Values of \(\alpha\) close to one retain the
sample spectrum, whereas smaller values of \(\alpha\) pull the eigenvalues more
strongly towards their grand mean. In this sense, shrinkage regularises the
relative eigenvalue structure of \(S\), while scalar corrections of the form
\(cS\) only rescale all eigenvalues by the same factor.


\subsection{Bayesian treatment of the shrinkage intensity}
\label{subsec:bayesian_alpha}

Our proposal is to treat \(\alpha\) as an auxiliary regularisation parameter.
Since \(0<\alpha<1\), we assign a proper prior density
\(\pi_\alpha(\alpha)\) on the unit interval. The framework is not tied to a
specific prior choice: different proper priors on \((0,1)\) can be used to encode
different assumptions about the amount of covariance regularisation. A natural and flexible class is
\[
\alpha\sim\mathrm{Beta}(a,b),
\]
with density
\[
\pi_\alpha(\alpha)
\propto
\alpha^{a-1}(1-\alpha)^{b-1}.
\]
In the simulations below, we use \(\alpha\sim\mathrm{Beta}(2,2)\) as a default
symmetric regularising prior. This prior assigns low density near the boundary
cases corresponding to the fully spherical target \((\alpha=0)\) and the
unregularised sample covariance \((\alpha=1)\), while allowing the likelihood to
concentrate near weak or strong shrinkage when supported by the data.

Let \(L_\alpha(x_o\mid\theta,S,n_s)\) denote the likelihood used for parameter
inference after replacing the covariance input by
\(\widehat{\Sigma}_\alpha\). In the Gaussian benchmark, this is the Gaussian
plug-in likelihood evaluated with covariance \(\widehat{\Sigma}_\alpha\).  In the main construction, this is the shrinkage-regularised Sellentin--Heavens-type
likelihood obtained by evaluating the Sellentin--Heavens functional form along
the covariance path \(\widehat{\Sigma}_\alpha\). Strictly speaking, this is not a
new exact Wishart--Jeffreys marginalisation, because
\(\widehat{\Sigma}_\alpha\) is not Wishart-distributed in general, so it should be
interpreted as a shrinkage-regularised Sellentin--Heavens-type working likelihood. The joint posterior is
\begin{equation}
\pi(\theta,\alpha\mid x_o,S)
\propto
L_\alpha(x_o\mid\theta,S,n_s)\,
\pi(\theta)\pi_\alpha(\alpha).
\label{eq:joint_posterior_alpha}
\end{equation}
The posterior distribution of the model parameters is obtained by marginalising
over the shrinkage intensity:
\begin{equation}
\pi(\theta\mid x_o,S)
=
\int_0^1
\pi(\theta,\alpha\mid x_o,S)\,d\alpha .
\label{eq:marginal_theta_alpha}
\end{equation}
Equivalently,
\[
\pi(\theta\mid x_o,S)
=
\int_0^1
\pi(\theta\mid x_o,S,\alpha)
\pi(\alpha\mid x_o,S)
\,d\alpha .
\]
Thus, the amount of covariance regularisation is not treated as fixed. The
likelihood induces a posterior distribution for \(\alpha\), and uncertainty about
this regularisation is integrated into the final posterior of \(\theta\).

\subsection{Posterior computation under a linear mean model}
\label{subsec:posterior_computation_alpha}

For nonlinear mean models or informative priors on \(\theta\), the joint
posterior in \eqref{eq:joint_posterior_alpha} can be sampled by standard Markov
chain Monte Carlo methods. In the simulation experiments considered in this
paper, however, we use a linear mean model,
\[
\mu_\theta=X\theta,
\qquad
k=\dim(\theta),
\]
together with a flat prior on \(\theta\). In this setting, conditional on
\(\alpha\), the posterior distribution of \(\theta\) is available analytically,
and the marginalisation over the one-dimensional shrinkage parameter can be
performed efficiently. For a fixed value of \(\alpha\), define
\[
\Omega_\alpha=\widehat{\Sigma}_\alpha^{-1},
\qquad
A_\alpha=X^\top\Omega_\alpha X,
\]
and assume that \(A_\alpha\) is nonsingular for the values of \(\alpha\) under
consideration. The weighted least-squares estimate and the corresponding
minimum quadratic form are
\[
\widehat{\theta}_\alpha
=
A_\alpha^{-1}X^\top\Omega_\alpha x_o,
\qquad
q_{\alpha}
=
(x_o-X\widehat{\theta}_\alpha)^\top
\Omega_\alpha
(x_o-X\widehat{\theta}_\alpha).
\]
The marginal likelihood of \(\alpha\), obtained by integrating out \(\theta\), is
available in closed form. For the Gaussian plug-in likelihood,
\begin{equation}
\log p_G(x_o\mid\alpha,S)
=
-\frac{1}{2}\log|\widehat{\Sigma}_\alpha|
-\frac{1}{2}\log|A_\alpha|
-\frac{1}{2}q_{\alpha}
+
\mathrm{const},
\label{eq:gaussian_alpha_marginal}
\end{equation}
where the constant does not depend on \(\alpha\). For the
shrinkage-regularised Sellentin--Heavens-type likelihood, the corresponding
marginal likelihood is
\begin{align}
&\log p_{SH}(x_o\mid\alpha,S)
=\\
&-\frac{1}{2}\log|\widehat{\Sigma}_\alpha|
-\frac{1}{2}\log|A_\alpha|
-\frac{n_s-k}{2}
\log
\left(
1+
\frac{q_{\alpha}}{n_s-1}
\right)
+
\mathrm{const},\nonumber
\label{eq:sh_alpha_marginal}
\end{align}
provided $n_s>k$. Combining either marginal likelihood with the prior on
\(\alpha\) gives
\begin{equation}
\pi(\alpha\mid x_o,S)
\propto
p(x_o\mid\alpha,S)\pi_\alpha(\alpha).
\label{eq:posterior_alpha}
\end{equation}
Thus, in the linear setting, the posterior distribution of \(\alpha\) can be
evaluated directly over \((0,1)\). We compute it on a fine grid
\(\{\alpha_m\}_{m=1}^{M}\) and normalise the resulting weights:
\begin{equation}
w_m
=
\frac{
p(x_o\mid\alpha_m,S)\pi_\alpha(\alpha_m)
}{
\sum_{\ell=1}^{M}
p(x_o\mid\alpha_\ell,S)\pi_\alpha(\alpha_\ell)
}.
\label{eq:alpha_weights}
\end{equation}
For the default \(\mathrm{Beta}(2,2)\) prior, these weights are proportional to
\(p(x_o\mid\alpha_m,S)\alpha_m(1-\alpha_m)\). For each fixed \(\alpha_m\), the conditional posterior
\(\pi(\theta\mid x_o,S,\alpha_m)\) is analytic: it is Gaussian under the
Gaussian plug-in likelihood and multivariate Student-\(t\), centred at
\(\widehat{\theta}_{\alpha_m}\), under the shrinkage-regularised
Sellentin--Heavens-type likelihood. The marginal posterior of \(\theta\) is
therefore approximated by the finite mixture
\begin{equation}
\pi(\theta\mid x_o,S)
\approx
\sum_{m=1}^{M}
w_m
\pi(\theta\mid x_o,S,\alpha_m).
\label{eq:theta_alpha_mixture}
\end{equation}
This mixture representation propagates uncertainty about the shrinkage intensity
into the posterior distribution of \(\theta\). 

Although the simulations use the linear setting in order to exploit these closed forms, the accompanying R and Python implementations also include an MCMC version for nonlinear mean models, including an exponential mean specification (see  Appendix \ref{app:examples}).

\section{Simulation evidence}
\label{sec:simulation}

We investigate the finite-sample behaviour of the proposed
covariance-shrinkage procedure through a set of Monte Carlo experiments. The
simulation evidence is organised by inferential objective. First, we verify the
scalar covariance-scaling results derived in Propositions~1 and~2. Second, we
examine how the posterior distribution of the shrinkage intensity reacts to the true
covariance structure, the likelihood and the number of simulations $n_s$ compared to the $p$. Third, we
evaluate whether covariance shrinkage improves posterior calibration and
inferential performance.

\subsection{Data-generating mechanism}
\label{subsec:dgp}

All simulations are based on the same data-generating mechanism. We consider the
linear mean model
\begin{equation}
y=X\theta_0+\varepsilon,
\qquad
\varepsilon\sim\mathcal{N}_p(0,\Sigma),
\label{eq:linear_dgp}
\end{equation}
where \(y\in\mathbb{R}^{p}\) is the observed data vector,
\(X\in\mathbb{R}^{p\times 2}\) is a fixed design matrix, and
\(\theta_0=(0,1)^\top\). The first column of \(X\) is equal to one, while the
second column contains values \(z_i\sim U(-1,1)\), generated once and kept fixed
within each scenario.

For each Monte Carlo replication \(r=1,\ldots,R\), we generate one observed data
vector \(y^{(r)}\). Independently, we generate \(n_s\) mock vectors from
\(\mathcal{N}_p(0,\Sigma)\) and compute the sample covariance matrix
\[
S^{(r)}
=
\frac{1}{n_s-1}
\sum_{j=1}^{n_s}
\left(x_j^{(r)}-\bar{x}^{(r)}\right)
\left(x_j^{(r)}-\bar{x}^{(r)}\right)^\top .
\]
Thus,
\[
(n_s-1)S^{(r)}\sim W_p(\Sigma,n_s-1),
\]
and \(S^{(r)}\) is independent of \(y^{(r)}\). 
We consider two covariance structures. The first is the identity covariance, $\Sigma=I_p$, while the second is an autoregressive covariance, $\Sigma_{ij}=\rho^{|i-j|}$, with $\rho=0.6$. The identity case represents a setting in which the spherical shrinkage target is
correctly specified, whereas the autoregressive case introduces structured
departures from sphericity. These two covariance structures therefore separate
the behaviour of the method when shrinkage towards \(\tau I_p\) is appropriate
from the behaviour obtained when the true covariance contains genuine
non-spherical dependence. The simulation grid varies both the data-vector dimension and the simulation
ratio:
\[
p\in\{20,50,100\},
\qquad
n_s/p\in\{2,5,10\}.
\]
The ratio \(n_s/p\) controls the severity of covariance-estimation noise. When
\(n_s/p\) is small, the sample covariance matrix is more variable, its eigenvalues
are more dispersed, and its inverse is more unstable. Larger values of \(n_s/p\)
represent settings in which the covariance estimate is better determined. The
grid therefore allows us to assess whether the proposed shrinkage formulation
responds both to the dimension of the data vector and to the amount of
simulation information available for covariance estimation.
\subsection{Scalar covariance corrections}
\label{subsec:sim_scalar}

\begin{figure}[!htb]
\centering
\includegraphics[width=\columnwidth]{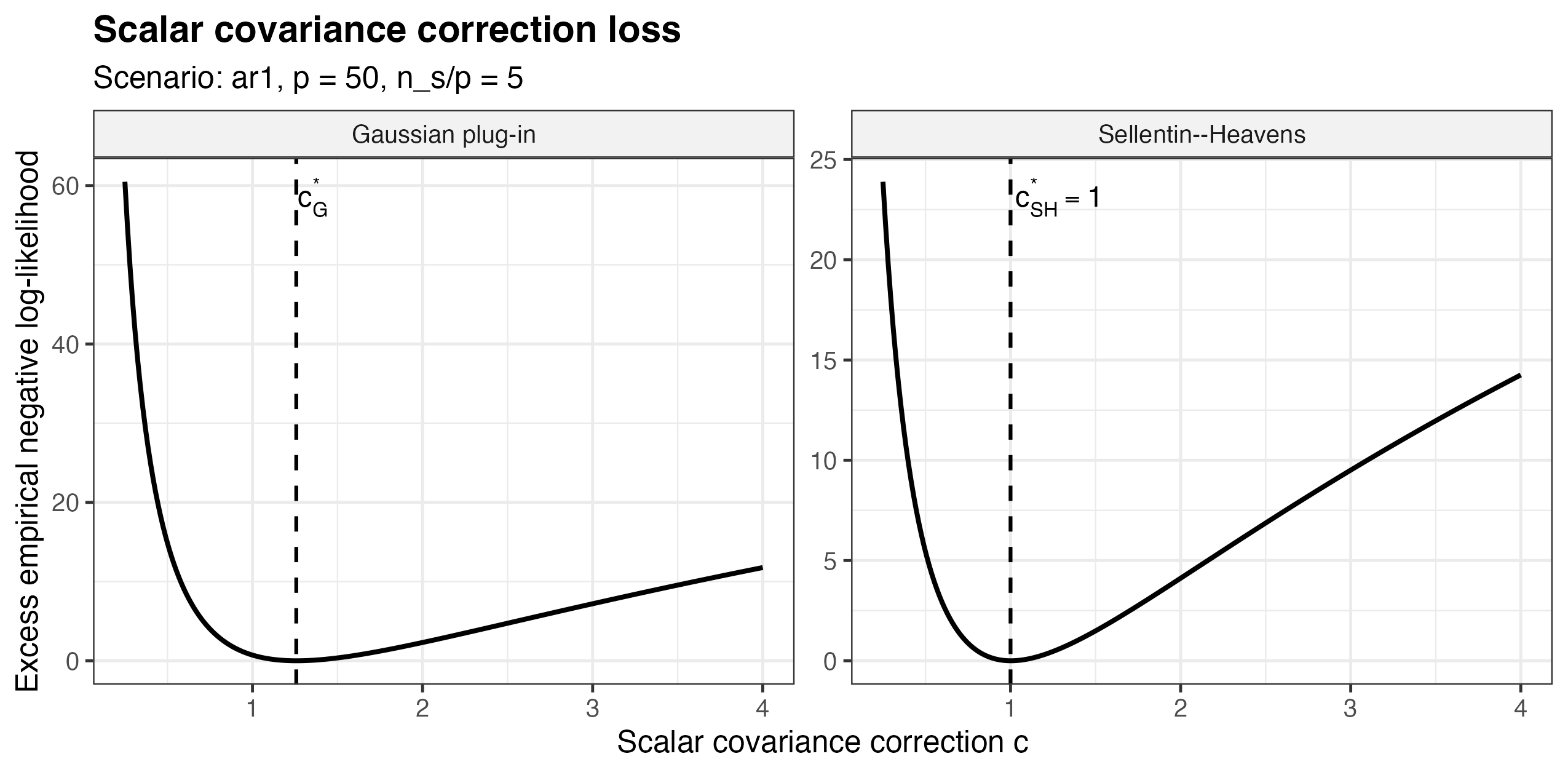}
\caption{Empirical excess negative log-likelihood loss for scalar covariance
corrections in a representative autoregressive scenario with \(p=50\) and
\(n_s/p=5\). The dashed lines indicate the theoretical scalar optima. The two
panels use separate vertical scales, and each loss is shifted by its minimum.
Under the Gaussian plug-in likelihood, the empirical minimum is close to the
Hartlap covariance-side scaling \(c_G^\star\). Under the Sellentin--Heavens
likelihood, the empirical minimum is close to \(c_{SH}^\star=1\).}
\label{fig:scalar_loss_curves}
\end{figure}

The first objective is to verify numerically the scalar correction results derived
in Propositions~1 and~2. For each replication, we compute the true residual vector
\[
e^{(r)}=y^{(r)}-X\theta_0 .
\]
By construction, \(e^{(r)}\sim\mathcal{N}_p(0,\Sigma)\), and \(e^{(r)}\) is
independent of \(S^{(r)}\). For a grid of scaling values \(c\), we evaluate the
empirical Gaussian negative log-likelihood loss
\[
\widehat{\mathcal{R}}_G(c)
=
\frac{1}{R}
\sum_{r=1}^{R}
\left[
\frac{1}{2}\log|cS^{(r)}|
+
\frac{1}{2}
e^{(r)\top}
(cS^{(r)})^{-1}
e^{(r)}
\right],
\]
and the empirical Sellentin--Heavens loss
\[
\widehat{\mathcal{R}}_{SH}(c)
=
\frac{1}{R}
\sum_{r=1}^{R}
\left[
\frac{1}{2}\log|cS^{(r)}|
+
\frac{n_s}{2}
\log
\left\{
1+
\frac{
e^{(r)\top}(cS^{(r)})^{-1}e^{(r)}
}
{n_s-1}
\right\}
\right].
\]
The empirical minimisers are compared with the theoretical scalar corrections
\[
c_G^\star=\frac{n_s-1}{n_s-p-2},
\qquad
c_{SH}^\star=1.
\]
Figure~\ref{fig:scalar_loss_curves} reports the empirical excess negative
log-likelihood loss for a representative autoregressive scenario with \(p=50\)
and \(n_s/p=5\). 
The results confirm the theoretical behaviour derived in
Propositions~1 and~2. Under the Gaussian plug-in likelihood, the loss is minimised
close to the Hartlap covariance-side scaling \(c_G^\star\). In contrast, under the
Sellentin--Heavens likelihood, the minimum is close to \(c_{SH}^\star=1\). 
Table~\ref{tab:scalar_corrections} provides the same numerical check across the
full simulation grid. 
\begin{table}[!htb]
\centering
\caption{Likelihood-loss optimal scalar covariance corrections. Theoretical
values are compared with empirical minimisers of the Monte Carlo loss curves.}
\label{tab:scalar_corrections}
\scriptsize
\setlength{\tabcolsep}{4pt}
\begin{tabular}{lrrrrrrr}
\hline
Cov. Type & \(p\) & \(n_s\) & \(n_s/p\) &
\(c_G^\star\) & \(\widehat c_G\) &
\(c_{SH}^\star\) & \(\widehat c_{SH}\) \\
\\
Identity & 20  & 40   & 2  & 2.167 & 2.042 & 1.000 & 0.958 \\
Identity & 20  & 100  & 5  & 1.269 & 1.229 & 1.000 & 0.973 \\
Identity & 20  & 200  & 10 & 1.118 & 1.108 & 1.000 & 0.988 \\
Identity & 50  & 100  & 2  & 2.062 & 2.057 & 1.000 & 1.003 \\
Identity & 50  & 250  & 5  & 1.258 & 1.259 & 1.000 & 1.003 \\
Identity & 50  & 500  & 10 & 1.114 & 1.108 & 1.000 & 1.003 \\
Identity & 100 & 200  & 2  & 2.031 & 2.057 & 1.000 & 1.018 \\
Identity & 100 & 500  & 5  & 1.254 & 1.259 & 1.000 & 1.003 \\
Identity & 100 & 1000 & 10 & 1.112 & 1.123 & 1.000 & 1.003 \\

AR(1) & 20  & 40   & 2  & 2.167 & 2.117 & 1.000 & 0.973 \\
AR(1) & 20  & 100  & 5  & 1.269 & 1.274 & 1.000 & 1.003 \\
AR(1) & 20  & 200  & 10 & 1.118 & 1.154 & 1.000 & 1.033 \\
AR(1) & 50  & 100  & 2  & 2.062 & 2.027 & 1.000 & 0.988 \\
AR(1) & 50  & 250  & 5  & 1.258 & 1.259 & 1.000 & 1.003 \\
AR(1) & 50  & 500  & 10 & 1.114 & 1.139 & 1.000 & 1.018 \\
AR(1) & 100 & 200  & 2  & 2.031 & 2.027 & 1.000 & 1.003 \\
AR(1) & 100 & 500  & 5  & 1.254 & 1.244 & 1.000 & 1.003 \\
AR(1) & 100 & 1000 & 10 & 1.112 & 1.123 & 1.000 & 1.003 \\
\hline
\end{tabular}
\vspace{2pt} \begin{minipage}{0.95\linewidth} 
\(c_G^\star=(n_s-1)/(n_s-p-2)\) denotes the Gaussian
likelihood-loss optimum, corresponding to the Hartlap covariance-side scaling.
\(c_{SH}^\star=1\) denotes the Sellentin--Heavens scalar optimum.
Empirical minimisers \(\widehat c_G\) and \(\widehat c_{SH}\) are obtained from
the Monte Carlo loss curves.
\end{minipage}
\end{table}
Across all dimensions, simulation ratios and covariance
structures, the empirical minimisers of the Gaussian loss are close to the
theoretical values \(c_G^\star=(n_s-1)/(n_s-p-2)\). The agreement is always strong but it increases as the simulation ratio increases, where the Gaussian optimum approaches
one and the empirical minimiser follows the same behaviour. By contrast, the
empirical minimisers of the Sellentin--Heavens loss remain close to
\(c_{SH}^\star=1\) in all scenarios. This also confirms that the difference between
Gaussian and Sellentin--Heavens scalar corrections is not driven by the particular
covariance structure used in the simulation, but by the likelihood itself. Moreover, these results show that
scalar covariance corrections are likelihood-dependent. The Hartlap-type covariance
inflation is appropriate for the Gaussian plug-in likelihood, whereas no
additional global rescaling is favoured once covariance uncertainty is propagated
through the Sellentin--Heavens likelihood. This provides justification for introducing shrinkage estimation.

\subsection{Posterior shrinkage intensity and conditioning}
\label{subsec:sim_alpha_conditioning}

The second objective is to study the behaviour of the posterior distribution of
the shrinkage intensity. Unlike \(c_G^\star\) and \(c_{SH}^\star\), the shrinkage
intensity does not have a true generative value in the simulation design. It is a
regularisation parameter controlling the compromise between the sample
covariance matrix and the spherical target. For each replication, we form
\[
\widehat{\Sigma}^{(r)}_\alpha
=
\alpha S^{(r)}
+
(1-\alpha)
\frac{\operatorname{tr}\{S^{(r)}\}}{p}I_p .
\]
Using the marginal likelihoods derived in
Section \ref{subsec:posterior_computation_alpha}, we compute the posterior density
of \(\alpha\) under both the Gaussian plug-in likelihood and the
Sellentin--Heavens likelihood with the default prior \(\alpha\sim\mathrm{Beta}(2,2)\). The posterior mean of
\(\alpha\) is used as a compact summary of the amount of likelihood-induced
shrinkage. We also compare the resulting covariance conditioning with that
obtained from the Ledoit--Wolf linear shrinkage estimator, which is used only as a matrix-estimation benchmark rather than as a posterior
marginalisation procedure.

Table~\ref{tab:alpha_summary} summarises the posterior mean of the shrinkage
intensity across all simulation scenarios. 
\begin{table}[!htb]
\centering
\caption{Posterior mean of the shrinkage intensity across simulation scenarios.
Values are averaged across Monte Carlo replications.}
\label{tab:alpha_summary}
\scriptsize
\setlength{\tabcolsep}{3pt}
\begin{tabular}{lrrrr}
\hline
Cov. Type & \(p\) & \(n_s/p\) & \(E_{SH}(\alpha)\) & \(E_G(\alpha)\)\\
\\
AR(1) & 20  & 2  & 0.630 & 0.570\\
AR(1) & 50  & 2  & 0.742 & 0.632\\
AR(1) & 100 & 2  & 0.796 & 0.659\\
AR(1) & 20  & 5  & 0.624 & 0.605\\
AR(1) & 50  & 5  & 0.750 & 0.714\\
AR(1) & 100 & 5  & 0.822 & 0.774\\
AR(1) & 20  & 10 & 0.632 & 0.624\\
AR(1) & 50  & 10 & 0.752 & 0.735\\
AR(1) & 100 & 10 & 0.830 & 0.809\\
Identity & 20  & 2  & 0.468 & 0.414\\
Identity & 50  & 2  & 0.407 & 0.318\\
Identity & 100 & 2  & 0.351 & 0.248\\
Identity & 20  & 5  & 0.473 & 0.461\\
Identity & 50  & 5  & 0.428 & 0.402\\
Identity & 100 & 5  & 0.375 & 0.338\\
Identity & 20  & 10 & 0.484 & 0.480\\
Identity & 50  & 10 & 0.455 & 0.446\\
Identity & 100 & 10 & 0.415 & 0.401\\
\hline
\end{tabular}
\vspace{2pt} \begin{minipage}{0.95\linewidth} 
\(E_{SH}(\alpha)\) denotes the posterior mean under the
Sellentin--Heavens likelihood, while \(E_G(\alpha)\) denotes the corresponding
posterior mean under the Gaussian plug-in likelihood. Smaller values of
\(\alpha\) indicate stronger shrinkage towards the spherical target.
 \end{minipage}
\end{table}
The posterior behaviour is coherent
with the spectral structure of the covariance matrix. In the identity-covariance
case, the posterior mean of \(\alpha\) is smaller, indicating stronger shrinkage
towards the spherical target. Under the Sellentin--Heavens likelihood, the
posterior mean ranges approximately from \(0.35\) to \(0.48\) in the identity
scenarios. This is expected because, when \(\Sigma=I_p\), the non-spherical
eigenvalue structure of the sample covariance matrix is mainly due to
finite-simulation noise. In the autoregressive covariance case, the posterior shifts towards larger values
of \(\alpha\). Under the Sellentin--Heavens likelihood, the posterior mean ranges
approximately from \(0.62\) to \(0.83\), indicating that the method retains more of
the sample covariance matrix when genuine non-spherical covariance structure is
present. 

Moreover, Table~\ref{tab:alpha_summary} shows that the Gaussian plug-in and
Sellentin--Heavens likelihoods induce different shrinkage behaviour. In most
scenarios, the posterior mean of \(\alpha\) is smaller under the Gaussian plug-in
likelihood than under the Sellentin--Heavens likelihood. Thus, Gaussian plug-in
inference tends to favour more aggressive shrinkage, whereas the heavier-tailed
Sellentin--Heavens likelihood allows more sample covariance structure to be
retained. This confirms that shrinkage is not only covariance-dependent, but also
likelihood-dependent.

The effect of shrinkage on numerical conditioning is reported in
Table~\ref{tab:condition_number}. 
\begin{table}[!htb]
\centering
\caption{Average condition number across simulation scenarios. Values are averaged
across Monte Carlo replications.}
\label{tab:condition_number}
\scriptsize
\setlength{\tabcolsep}{2.7pt}
\begin{tabular}{lrrrrr}
\hline
Cov. Type & \(p\) & \(n_s/p\) & \(\kappa(S)\) & \(\kappa_{SH}\) & \(\kappa_{LW}\)\\
\\
AR(1) & 20  & 2  & 109.0 & 8.9  & 9.6\\
AR(1) & 50  & 2  & 131.9 & 16.3 & 11.4\\
AR(1) & 100 & 2  & 144.3 & 22.0 & 12.3\\
AR(1) & 20  & 5  & 32.7  & 6.8  & 12.9\\
AR(1) & 50  & 5  & 38.8  & 11.7 & 15.6\\
AR(1) & 100 & 5  & 41.7  & 15.9 & 16.7\\
AR(1) & 20  & 10 & 23.0  & 6.2  & 14.6\\
AR(1) & 50  & 10 & 26.4  & 9.9  & 17.1\\
AR(1) & 100 & 10 & 27.9  & 13.2 & 18.1\\
\addlinespace
Identity & 20  & 2  & 25.6 & 3.1 & 1.2\\
Identity & 50  & 2  & 29.2 & 2.8 & 1.1\\
Identity & 100 & 2  & 30.9 & 2.5 & 1.0\\
Identity & 20  & 5  & 5.8  & 2.1 & 1.1\\
Identity & 50  & 5  & 6.2  & 2.0 & 1.0\\
Identity & 100 & 5  & 6.5  & 1.9 & 1.0\\
Identity & 20  & 10 & 3.3  & 1.7 & 1.0\\
Identity & 50  & 10 & 3.5  & 1.7 & 1.0\\
Identity & 100 & 10 & 3.6  & 1.7 & 1.0\\
\hline
\end{tabular}
\vspace{2pt} \begin{minipage}{0.95\linewidth} 
\(\kappa(S)\) is the condition number of the sample covariance matrix,
\(\kappa_{SH}\) is the condition number of the posterior-mean proposed shrinkage
covariance under the Sellentin--Heavens likelihood, and \(\kappa_{LW}\) is the
condition number of the Ledoit--Wolf covariance estimator.
 \end{minipage}
\end{table}
In all scenarios, replacing the sample covariance
matrix \(S\) by a shrinkage covariance substantially reduces the condition number.
The reduction is particularly pronounced in the autoregressive scenarios, where
the sample covariance matrix is poorly conditioned when \(n_s/p\) is small. For
example, when \(p=100\) and \(n_s/p=2\), the average condition number decreases from
approximately \(144\) for the sample covariance matrix to approximately \(22\) for
the proposed shrinkage covariance under the Sellentin--Heavens likelihood. Ledoit--Wolf shrinkage often gives the strongest numerical regularisation,
especially in the identity-covariance case, where the covariance is close to the
spherical target. In the autoregressive scenarios, the proposed shrinkage
covariance is generally less aggressive than Ledoit--Wolf shrinkage, retaining more
of the sample covariance structure while still providing a substantial improvement
in conditioning. This distinction is important because the objective of the
present method is not only to improve matrix conditioning, but to induce
likelihood-dependent regularisation and propagate uncertainty in the amount of
shrinkage.

\subsection{Posterior calibration and inferential performance}
\label{subsec:sim_calibration}

The final objective is to evaluate the inferential consequences of covariance
shrinkage. For each replication, posterior inference is carried out for
\(\theta=(\theta_0,\theta_1)^\top\). We compare five classes of procedures: the oracle Gaussian posterior using the true
covariance matrix \(\Sigma\); Gaussian plug-in procedures based on \(S\); the unregularised Sellentin--Heavens posterior
based on \(S\); Sellentin--Heavens plug-in shrinkage using either Ledoit--Wolf; and the proposed Sellentin--Heavens posterior with
\(\alpha\) marginalised. The Gaussian procedures are included to separate the
effect of covariance shrinkage from the effect of covariance marginalisation in
the Sellentin--Heavens likelihood. For each method, we compute posterior means, posterior standard deviations,
marginal credible intervals and joint credible regions. The main calibration
summary reported below is the empirical coverage of 95\% joint credible regions,
together with the mean log score at the true parameter value and the average
posterior area.

Table~\ref{tab:simII_joint_cov_ratio} summarises the joint posterior calibration
results separately for the two covariance structures and for the three simulation
ratios \(n_s/p\). 
\begin{table*}[!htb]
\centering
\caption{Joint posterior performance by covariance structure and simulation ratio.
Values are averaged over \(p\in\{20,50,100\}\).}
\label{tab:simII_joint_cov_ratio}
\scriptsize
\setlength{\tabcolsep}{4pt}
\begin{tabular}{llrrrrrrrrr}
\hline
& & \multicolumn{3}{c}{Joint coverage} &
\multicolumn{3}{c}{Log score} &
\multicolumn{3}{c}{Area} \\
\cline{3-5}\cline{6-8}\cline{9-11}
Cov. Type & Method
& \(2\) & \(5\) & \(10\)
& \(2\) & \(5\) & \(10\)
& \(2\) & \(5\) & \(10\) \\
\hline
AR(1) & Oracle Gaussian
& 0.954 & 0.953 & 0.942
& -1.020 & -1.022 & -1.059
& 0.186 & 0.186 & 0.186 \\
AR(1) & Gaussian \(S\)
& 0.541 & 0.849 & 0.907
& -3.342 & -1.392 & -1.177
& 0.096 & 0.151 & 0.168 \\
AR(1) & SH \(S\)
& 0.802 & 0.906 & 0.927
& -1.996 & -1.297 & -1.156
& 0.197 & 0.190 & 0.188 \\
AR(1) & SH Ledoit--Wolf
& 0.913 & 0.915 & 0.922
& -1.239 & -1.178 & -1.132
& 0.181 & 0.180 & 0.181 \\
AR(1) & SH marginal \(\alpha\)
& 0.907 & 0.913 & 0.911
& -1.282 & -1.164 & -1.128
& 0.181 & 0.166 & 0.161 \\
Identity & Oracle Gaussian
& 0.954 & 0.948 & 0.954
& 0.459 & 0.463 & 0.460
& 0.044 & 0.044 & 0.044 \\
Identity & Gaussian \(S\)
& 0.561 & 0.853 & 0.920
& -1.734 & 0.155 & 0.328
& 0.023 & 0.036 & 0.040 \\
Identity & SH \(S\)
& 0.798 & 0.916 & 0.941
& -0.441 & 0.246 & 0.349
& 0.047 & 0.045 & 0.045 \\
Identity & SH Ledoit--Wolf
& 0.991 & 0.969 & 0.965
& 0.383 & 0.448 & 0.458
& 0.069 & 0.054 & 0.049 \\
Identity & SH marginal \(\alpha\)
& 0.983 & 0.962 & 0.962
& 0.363 & 0.430 & 0.440
& 0.065 & 0.052 & 0.048 \\
\hline
\end{tabular}
\vspace{2pt} \begin{minipage}{0.95\linewidth} 
The columns labelled \(2\), \(5\) and \(10\) refer to the simulation ratio
\(n_s/p\). Joint coverage denotes the empirical coverage of nominal 95\% joint
credible regions. Log score is evaluated at the true parameter value. Area
denotes the average area of the joint credible region.
 \end{minipage}
\end{table*}
This representation preserves the role of the simulation ratio,
which controls the severity of covariance-estimation noise. The strongest
differences between methods occur when \(n_s/p=2\), that is, when the covariance
matrix is estimated from relatively few simulations per data-vector component. In both covariance structures, the Gaussian plug-in posterior based on the raw
sample covariance (S) shows severe undercoverage when \(n_s/p\) is small. For
\(n_s/p=2\), joint coverage is 0.541 in the autoregressive case and 0.561 in the
identity case. This undercoverage is accompanied by smaller posterior regions
than the oracle posterior, indicating that the Gaussian plug-in likelihood
underestimates posterior uncertainty when covariance noise is not accounted for.
As the simulation ratio increases, the sample covariance becomes better
determined and the Gaussian plug-in posterior improves, reaching joint coverage
above 0.90 for \(n_s/p=10\) in both covariance structures.

Replacing the Gaussian plug-in likelihood by the Sellentin--Heavens likelihood
substantially improves calibration. In the autoregressive case, joint coverage
increases from 0.541 to 0.802 when \(n_s/p=2\), and from 0.849 to 0.906 when
\(n_s/p=5\). A similar pattern is observed in the identity case, where coverage
increases from 0.561 to 0.798 for \(n_s/p=2\), and from 0.853 to 0.916 for
\(n_s/p=5\). These results confirm that covariance marginalisation mitigates the
inferential consequences of using a covariance matrix estimated from a finite
number of simulations. However, the unregularised Sellentin--Heavens posterior
still remains below nominal coverage in the most difficult setting
\(n_s/p=2\), showing that covariance marginalisation alone does not fully remove
the effect of noisy covariance eigenvalues.

The shrinkage-regularised Sellentin--Heavens procedures further improve
performance, especially when the simulation ratio is small. In the autoregressive
case with \(n_s/p=2\), Ledoit--Wolf shrinkage and marginalisation over \(\alpha\)
increase joint coverage to 0.913 and 0.907, respectively, and substantially
improve the log score relative to the unregularised Sellentin--Heavens posterior.
For larger simulation ratios, the difference between unregularised and
shrinkage-regularised procedures becomes smaller, as expected, because the sample
covariance is less noisy.

The identity case highlights the role of the shrinkage target. Since the true
covariance is spherical, shrinkage towards \(\tau I_p\) is well aligned with the
data-generating covariance. The shrinkage-regularised procedures therefore give
very high coverage, particularly when \(n_s/p=2\). This should not be interpreted
as superiority over the oracle Gaussian posterior. The oracle posterior is
already correctly calibrated, with coverage close to 0.95 and smaller credible
regions. The higher coverage of the shrinkage procedures in the identity case is
a conservative effect, reflected by their larger posterior areas.

These results suggest that the Ledoit--Wolf estimator is a strong plug-in
benchmark and gives performance comparable to the proposed approach in this
simulation design. This is expected, since the simulations use a spherical
shrinkage target, for which Ledoit--Wolf-type shrinkage is particularly well
suited. At comparable empirical performance, however, our proposed approach is more
informative and flexible from an inferential perspective: it learns the shrinkage intensity through the likelihood, propagates its uncertainty into the posterior of the scientific parameters, and can be extended directly to non-spherical targets, including analytical, surrogate-based or survey-informed covariance
models. 

\section{Conclusions}
\label{sec:conclusion}

This paper studied covariance regularisation for simulation-based likelihood
inference when the covariance matrix is estimated from a finite number of mock
data vectors. The starting point was the distinction between precision-matrix
debiasing and covariance-marginalised likelihood inference. Hartlap scaling
corrects the expectation of the inverse sample covariance while retaining a
Gaussian plug-in likelihood. The Sellentin--Heavens likelihood instead modifies
the likelihood itself by marginalising over covariance uncertainty. These two
corrections therefore address related but distinct inferential problems.

We first showed that scalar covariance corrections are likelihood-dependent.
Under a Gaussian working likelihood, the expected negative log-likelihood is
minimised by the Hartlap covariance-side scaling,
\[
c_G^\star=\frac{n_s-1}{n_s-p-2}.
\]
Under the Sellentin--Heavens likelihood, the corresponding scalar optimum is
\[
c_{SH}^\star=1.
\]
The contribution of this result is interpretative rather than algebraic: Hartlap
scaling is recovered as the scalar correction preferred by Gaussian plug-in
inference, but it should not be transferred automatically to
covariance-marginalised likelihoods. Once covariance uncertainty has been
incorporated through the Sellentin--Heavens likelihood, an additional global
rescaling of \(S\) is not favoured within the scalar class \(\{cS:c>0\}\).

This does not imply that the unregularised sample covariance is optimal among
richer covariance estimators. Scalar scaling changes only the global covariance
amplitude, whereas shrinkage regularises the noisy eigenvalue structure of the
sample covariance matrix. We therefore introduced a shrinkage-intensity
formulation in which
\[
\widehat{\Sigma}_\alpha
=
\alpha S+(1-\alpha)\tau I_p
\]
is used as the regularised covariance input inside the likelihood. The shrinkage
intensity \(\alpha\) is treated as an auxiliary regularisation parameter. A prior
is assigned to $\alpha$, the likelihood induces its posterior distribution, and it is
marginalised in the final posterior of the scientific parameters.

The simulation results support the central claim of the paper: covariance
marginalisation and shrinkage regularisation are complementary. The
Sellentin--Heavens likelihood improves calibration relative to the Gaussian
plug-in likelihood by accounting for finite-simulation covariance uncertainty,
whereas shrinkage further stabilises the noisy covariance structure entering the
likelihood. Combining the two therefore addresses limitations that remain when
either covariance marginalisation or shrinkage regularisation is used in
isolation.

In this paper, we used the simple spherical target \(\tau I_p\) in order to
isolate the role of the shrinkage intensity. This choice also gives a favourable
setting for standard Ledoit--Wolf-type shrinkage, which explains why the
Ledoit--Wolf benchmark performs very strongly in the simulations. The proposed
framework is not intended to replace Ledoit--Wolf as a covariance estimator under
a fixed matrix loss. Its aim is instead to provide an inferential treatment of
shrinkage uncertainty inside a covariance-marginalised likelihood. A natural extension is to replace the spherical target by a structured
covariance target,
\[
\widehat{\Sigma}_{\alpha,T}
=
\alpha S+(1-\alpha)T .
\]
The target \(T\) could be an analytical covariance model, a covariance obtained
from surrogate-assisted methods, or a validated survey covariance matrix. Within
the proposed framework, this extension is direct: the same prior and posterior
marginalisation over \(\alpha\) can be used to quantify how strongly inference
should rely on the noisy sample covariance relative to the chosen structured
target. By contrast, plug-in shrinkage estimators such as Ledoit--Wolf require
the shrinkage intensity to be derived or estimated for the specific target and
matrix-estimation criterion. Target-informed shrinkage within the proposed
likelihood framework is therefore a promising direction for future work.

\bibliographystyle{plainnat}
\bibliography{sample}

\appendix
\section{Examples with simulated data}
\label{app:examples}

This section provides two simple examples illustrating how the proposed covariance-shrinkage procedure can be applied in practice. The examples
are not intended as additional simulation studies. Their purpose is to show the
computational distinction between the linear case, where the mixture
representation in \eqref{eq:theta_alpha_mixture} can be used directly, and a
nonlinear case, where the joint posterior in \eqref{eq:joint_posterior_alpha}
is sampled by MCMC. The code used for the examples, including the R and Python
implementations of the linear method and the Stan examples for nonlinear mean
models, is available in the repository
\url{https://github.com/raffmattera/cosmo-shrinkage-inference}.

\subsection{Linear example}
\label{app:example_linear}

The first example uses the same linear data-generating structure described in
Section~\ref{subsec:dgp}. We set \(\theta_0=0\), \(\theta_1=1\), use a regular grid
\(z_i\in[-1,1]\) for the second column of the design matrix, and generate the
observed data vector from the linear model in Eq.~\eqref{eq:linear_dgp}. The
covariance matrix is autoregressive, with \(\Sigma_{ij}=\rho^{|i-j|}\) and
\(\rho=0.7\). Independently generated mock data vectors are used to compute the
sample covariance matrix \(S\).

The shrinkage covariance estimator is the linear family in
Eq.~\eqref{eq:shrinkage_estimator}, with the default prior
\(\alpha\sim\operatorname{Beta}(2,2)\). Since the mean model is linear, the
conditional posterior of \(\theta\) given \(\alpha\) is available in closed form,
and the posterior of \(\alpha\) is evaluated by one-dimensional quadrature as in
Section~\ref{subsec:posterior_computation_alpha}. The final posterior of
\(\theta\) is then obtained from the finite-mixture approximation in
Eq.~\eqref{eq:theta_alpha_mixture}.

Table~\ref{tab:example_linear} reports the posterior summaries obtained under
the Sellentin--Heavens and Gaussian plug-in likelihoods. In this specific
realisation, the posterior distribution of \(\alpha\) is concentrated towards high
values, indicating that the likelihood favours retaining most of the sample
covariance structure while still applying a non-negligible amount of shrinkage.
The Sellentin--Heavens posterior produces slightly wider credible intervals than
the Gaussian plug-in likelihood, reflecting the additional uncertainty associated
with the simulation-estimated covariance matrix.
\begin{table}[!htb]
\centering
\caption{Posterior summaries for the linear example with simulated data.}
\label{tab:example_linear}
\scriptsize
\setlength{\tabcolsep}{4pt}
\begin{tabular}{llrrrrr}
\hline
Likelihood & Parameter & True & Mean & SD & 2.5\% & 97.5\% \\
\\
SH & \(\theta_0\) & 0.000 & -0.034 & 0.319 & -0.668 & 0.577 \\
SH & \(\theta_1\) & 1.000 &  1.932 & 0.535 &  0.883 & 2.982 \\
SH & \(\alpha\)   & --    &  0.879 & --    &  0.668 & 0.984 \\
\hline
Gaussian & \(\theta_0\) & 0.000 & -0.040 & 0.290 & -0.608 & 0.531 \\
Gaussian & \(\theta_1\) & 1.000 &  1.940 & 0.477 &  1.025 & 2.875 \\
Gaussian & \(\alpha\)   & --    &  0.858 & --    &  0.634 & 0.976 \\
\hline
\end{tabular}
\vspace{2pt} \begin{minipage}{0.95\linewidth} 
SH denotes the Sellentin--Heavens covariance-marginalised likelihood.
The shrinkage intensity \(\alpha\) has no
data-generating value.
 \end{minipage}
\end{table}
This example illustrates the computational advantage of the analytic mixture
representation. Once \(S\) and \(X\) are specified, posterior inference requires
only the evaluation of the marginal likelihood of \(\alpha\) over a
one-dimensional grid and posterior simulation from the corresponding conditional
Gaussian or Student-\(t\) distributions.

\subsection{Nonlinear example}
\label{app:example_nonlinear}

The second example uses the nonlinear mean function
\[
\mu_i(\theta)=\exp(\theta_0+\theta_1 z_i),
\]
with Gaussian errors and an autoregressive covariance matrix. The same mock-based
covariance-estimation mechanism used in Section~\ref{subsec:dgp} is retained: a
single observed data vector is generated from the nonlinear model, while an
independent set of mock data vectors is used to compute \(S\). The shrinkage
covariance estimator is again given by \eqref{eq:shrinkage_estimator} and is
used inside the Sellentin--Heavens likelihood.

The computational difference from the linear example is that the residual vector
\(r(\theta)=y-\mu(\theta)\) is no longer linear in \(\theta\). As a result, the
quadratic form entering the likelihood is not quadratic in \(\theta\), and the
conditional posterior of \(\theta\) given \(\alpha\) is no longer available in
closed form. Therefore, the analytic mixture representation in
\eqref{eq:theta_alpha_mixture} cannot be used directly. Instead, we sample the
joint posterior in \eqref{eq:joint_posterior_alpha} using Stan, assigning the
same \(\operatorname{Beta}(2,2)\) prior to \(\alpha\) and weakly informative
Gaussian priors to \(\theta_0\) and \(\theta_1\).

\begin{figure}[!htb]
\centering
\includegraphics[width=\columnwidth]{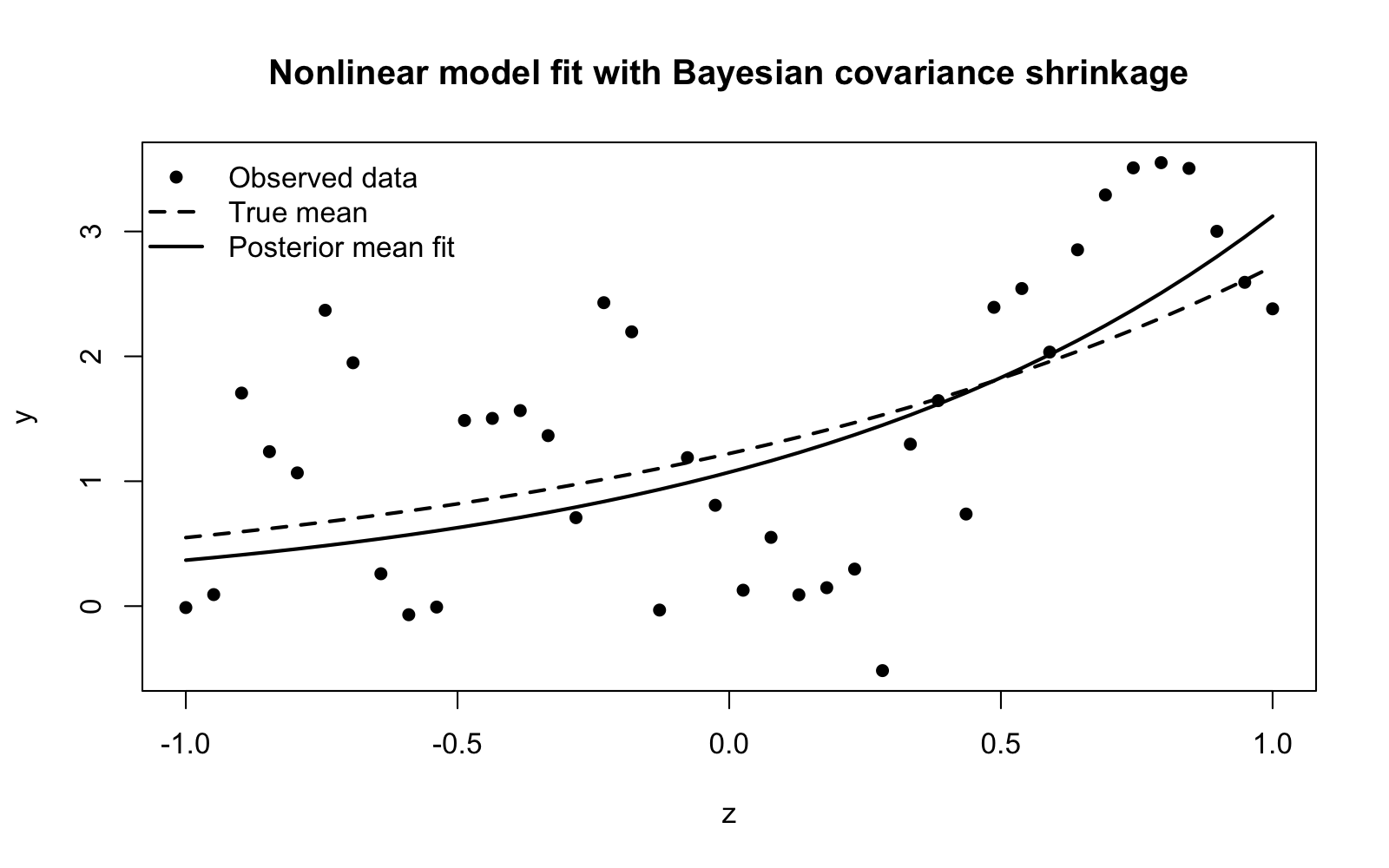}
\caption{Nonlinear example with simulated data. Points denote the observed data
vector, the dashed curve is the true mean function
\(\mu_i(\theta)=\exp(\theta_0+\theta_1 z_i)\), and the solid curve is the posterior
mean fit obtained by sampling the joint posterior of \((\theta,\alpha)\) with
Stan.}
\label{fig:example_nonlinear_fit}
\end{figure}

Table~\ref{tab:example_nonlinear} reports the posterior summaries for the
nonlinear example. The true values of the nonlinear parameters are contained in
the posterior credible intervals. The posterior mean of \(\alpha\) is
approximately 0.68, indicating moderate shrinkage towards the spherical target
while retaining a substantial part of the sample covariance structure. The
effective sample sizes and \(\hat{R}\) diagnostics from Stan indicate stable
sampling in this low-dimensional example.
\begin{table}[!htb]
\centering
\caption{Posterior summaries for the nonlinear example fitted by Stan.}
\label{tab:example_nonlinear}
\scriptsize
\setlength{\tabcolsep}{4pt}
\begin{tabular}{lrrrrrr}
\hline
Parameter & True & Mean & SD & 2.5\% & 50\% & 97.5\% \\
\\
\(\theta_0\) & 0.200 & 0.070 & 0.340 & -0.760 & 0.130 & 0.560 \\
\(\theta_1\) & 0.800 & 1.070 & 0.440 &  0.370 & 1.010 & 2.110 \\
\(\alpha\)   & --    & 0.680 & 0.170 &  0.290 & 0.710 & 0.940 \\
\hline
\end{tabular}
\vspace{2pt} \begin{minipage}{0.95\linewidth} 
The nonlinear mean is \(\mu_i(\theta)=\exp(\theta_0+\theta_1 z_i)\).
The shrinkage intensity \(\alpha\) has no
data-generating value.
 \end{minipage}
\end{table}
Figure~\ref{fig:example_nonlinear_fit} shows the observed data, the true
nonlinear mean function and the posterior mean fit. The posterior mean curve
captures the increasing nonlinear trend of the data, although the fit is
naturally affected by the covariance structure and by the finite number of
observations. The point of the example is not to introduce a new simulation
benchmark, but to show that the same shrinkage hierarchy can be implemented when
the mean model is nonlinear and the analytic conditional posterior is unavailable.

\section{Technical results}
\label{app:multivariate_results}

We collect here the standard multivariate distributional results used in the
proofs. Let \(W\sim W_p(\Sigma,\nu)\) denote a \(p\)-dimensional Wishart random
matrix with scale matrix \(\Sigma\) and \(\nu\) degrees of freedom. First, if \(x_1,\ldots,x_{n_s}\) are independent Gaussian mock data vectors with
covariance matrix \(\Sigma\), and
\[
S
=
\frac{1}{n_s-1}
\sum_{i=1}^{n_s}
(x_i-\bar{x})(x_i-\bar{x})^\top,
\]
then
\begin{equation}
(n_s-1)S\sim W_p(\Sigma,n_s-1).
\label{eq:app_wishart_sample_cov}
\end{equation}
Second, if \(W\sim W_p(\Sigma,\nu)\) with \(\nu>p+1\), then the inverse moment is
\begin{equation}
E(W^{-1})
=
\frac{1}{\nu-p-1}\Sigma^{-1}.
\label{eq:app_inverse_wishart_moment}
\end{equation}
Consequently, for \(W=(n_s-1)S\) and \(\nu=n_s-1\),
\begin{equation}
E(S^{-1})
=
\frac{n_s-1}{n_s-p-2}\Sigma^{-1},
\qquad n_s>p+2.
\label{eq:app_inverse_sample_cov_moment}
\end{equation}
Third, if \(x_o\sim \mathcal{N}_p(\mu,\Sigma)\) is independent of \(S\), then
\[
E\{(x_o-\mu)(x_o-\mu)^\top\}=\Sigma,
\]
and, for any random matrix \(M\) independent of \(x_o\),
\begin{equation}
E\left[
(x_o-\mu)^\top M(x_o-\mu)
\right]
=
E\left[
\operatorname{tr}(M\Sigma)
\right].
\label{eq:app_trace_identity}
\end{equation}
Fourth, let \(W\sim W_p(\Sigma,\nu)\), let
\[
A=\Sigma^{-1/2}W\Sigma^{-1/2},
\]
and let
\[
z=\Sigma^{-1/2}(x_o-\mu)\sim\mathcal{N}_p(0,I_p)
\]
be independent of \(A\). Then
\[
A\sim W_p(I_p,\nu),
\]
and the quadratic form \(z^\top A^{-1}z\) admits the Hotelling--beta
representation
\begin{equation}
z^\top A^{-1}z
\overset{d}{=}
\frac{B}{1-B},
\qquad
B\sim
\mathrm{Beta}
\left(
\frac{p}{2},
\frac{\nu+1-p}{2}
\right).
\label{eq:app_hotelling_beta}
\end{equation}
In the present paper, \(\nu=n_s-1\), so
\begin{equation}
z^\top A^{-1}z
\overset{d}{=}
\frac{B}{1-B},
\qquad
B\sim
\mathrm{Beta}
\left(
\frac{p}{2},
\frac{n_s-p}{2}
\right).
\label{eq:app_hotelling_beta_ns}
\end{equation}
In particular,
\begin{equation}
E(B)
=
\frac{p}{n_s}.
\label{eq:app_beta_mean}
\end{equation}
These results are standard consequences of the Gaussian--Wishart model
\citep[e.g.][]{muirhead2009aspects}.

\section{Proofs}
\label{app:proofs}

\subsection{Proof of Proposition 1}
\label{app:proof_prop1}

We evaluate the scalar covariance rule
\[
\widehat{\Sigma}_{c}=cS,
\qquad c>0,
\]
under the expected Gaussian negative log-likelihood loss. The expectation is taken
with respect to both the observed data vector \(x_o\) and the simulation-based
covariance estimator \(S\). Under the assumptions of the proposition,
\[
x_o\sim\mathcal{N}_p(\mu,\Sigma),
\qquad
(n_s-1)S\sim W_p(\Sigma,n_s-1),
\]
with \(x_o\) and \(S\) independent. Up to constants independent of \(c\), the Gaussian negative log-likelihood based
on the covariance estimator \(cS\) is
\[
\ell_G(c)
=
\frac{1}{2}\log|cS|
+
\frac{1}{2}
(x_o-\mu)^\top(cS)^{-1}(x_o-\mu).
\]
The corresponding risk function is
\[
\mathcal{R}_G(c)
=
E\{\ell_G(c)\}.
\]
Using
\[
\log|cS|=p\log c+\log|S|,
\qquad
(cS)^{-1}=c^{-1}S^{-1},
\]
we obtain
\begin{equation}
\mathcal{R}_G(c)
=
\frac{p}{2}\log c
+
\frac{1}{2c}
E\left[
(x_o-\mu)^\top S^{-1}(x_o-\mu)
\right]
+
\mathrm{const}.
\label{eq:app_rg_first}
\end{equation}
By \eqref{eq:app_trace_identity},
\[
E\left[
(x_o-\mu)^\top S^{-1}(x_o-\mu)
\right]
=
E\left[
\operatorname{tr}(S^{-1}\Sigma)
\right].
\]
Using \eqref{eq:app_inverse_sample_cov_moment},
\[
E\left[
\operatorname{tr}(S^{-1}\Sigma)
\right]
=
\operatorname{tr}\{E(S^{-1})\Sigma\}
=
\frac{n_s-1}{n_s-p-2}\operatorname{tr}(I_p)
=
\frac{(n_s-1)p}{n_s-p-2}.
\]
Therefore,
\begin{equation}
\mathcal{R}_G(c)
=
\frac{1}{2}
\left[
p\log c
+
\frac{1}{c}
\frac{(n_s-1)p}{n_s-p-2}
\right]
+
\mathrm{const}.
\label{eq:app_rg_closed}
\end{equation}
Differentiating with respect to \(c\) gives
\[
\frac{d\mathcal{R}_G(c)}{dc}
=
\frac{p}{2c}
-
\frac{1}{2c^2}
\frac{(n_s-1)p}{n_s-p-2}.
\]
Setting the derivative equal to zero yields
\begin{equation}
c_G^\star
=
\frac{n_s-1}{n_s-p-2}.
\label{eq:app_cg_star}
\end{equation}
Moreover,
\[
\frac{d^2\mathcal{R}_G(c)}{dc^2}
=
-\frac{p}{2c^2}
+
\frac{1}{c^3}
\frac{(n_s-1)p}{n_s-p-2},
\]
and, evaluated at \(c=c_G^\star\),
\[
\left.
\frac{d^2\mathcal{R}_G(c)}{dc^2}
\right|_{c=c_G^\star}
=
\frac{p}{2(c_G^\star)^2}>0.
\]
Thus \(c_G^\star\) is the unique minimiser of the expected Gaussian negative
log-likelihood loss. Finally,
\[
(c_G^\star S)^{-1}
=
\frac{n_s-p-2}{n_s-1}S^{-1}.
\]
Hence the likelihood-loss optimal covariance scaling is equivalent, on the
precision side, to the Hartlap-corrected inverse covariance estimator.

\subsection{Proof of Proposition 2}
\label{app:proof_prop2}

We now evaluate the same scalar covariance rule
\[
\widehat{\Sigma}_{c}=cS,
\qquad c>0,
\]
under the expected negative log-likelihood induced by the Sellentin--Heavens
likelihood. Again, the expectation is taken with respect to both \(x_o\) and \(S\).
Under the assumptions of the proposition,
\[
x_o\sim\mathcal{N}_p(\mu,\Sigma),
\qquad
(n_s-1)S\sim W_p(\Sigma,n_s-1),
\]
with \(x_o\) and \(S\) independent. Let
\[
q
=
(x_o-\mu)^\top S^{-1}(x_o-\mu).
\]
The Sellentin--Heavens likelihood based on the scaled covariance estimator \(cS\)
contains the quadratic term
\[
q_c
=
(x_o-\mu)^\top(cS)^{-1}(x_o-\mu)
=
\frac{q}{c}.
\]
Up to constants independent of \(c\), the negative log-likelihood is therefore
\[
\ell_{SH}(c)
=
\frac{1}{2}\log|cS|
+
\frac{n_s}{2}
\log
\left(
1+\frac{q}{c(n_s-1)}
\right).
\]
The corresponding risk function is
\[
\mathcal{R}_{SH}(c)
=
E\{\ell_{SH}(c)\}.
\]
Using \(\log|cS|=p\log c+\log|S|\), we obtain
\begin{equation}
\mathcal{R}_{SH}(c)
=
\frac{p}{2}\log c
+
\frac{n_s}{2}
E\left[
\log
\left(
1+
\frac{q}{c(n_s-1)}
\right)
\right]
+
\mathrm{const}.
\label{eq:app_rsh_first}
\end{equation}
It remains to express the distribution of \(q/(n_s-1)\). Let
\[
W=(n_s-1)S,
\qquad
z=\Sigma^{-1/2}(x_o-\mu),
\qquad
A=\Sigma^{-1/2}W\Sigma^{-1/2}.
\]
Then
\[
z\sim\mathcal{N}_p(0,I_p),
\qquad
A\sim W_p(I_p,n_s-1),
\]
and \(z\) and \(A\) are independent. Moreover,
\[
q
=
(x_o-\mu)^\top S^{-1}(x_o-\mu)
=
(n_s-1)z^\top A^{-1}z.
\]
Therefore,
\[
\frac{q}{n_s-1}
=
z^\top A^{-1}z.
\]
By \eqref{eq:app_hotelling_beta_ns},
\[
\frac{q}{n_s-1}
\overset{d}{=}
\frac{B}{1-B},
\qquad
B\sim
\mathrm{Beta}
\left(
\frac{p}{2},
\frac{n_s-p}{2}
\right).
\]
Substituting this representation into \eqref{eq:app_rsh_first} gives, up to
constants independent of \(c\),
\begin{equation}
\mathcal{R}_{SH}(c)
=
\frac{p}{2}\log c
+
\frac{n_s}{2}
E_B
\left[
\log
\left(
1+
\frac{B}{c(1-B)}
\right)
\right]
+
\mathrm{const}.
\label{eq:app_rsh_beta}
\end{equation}
Differentiating with respect to \(c\) gives
\begin{equation}
\frac{\partial \mathcal{R}_{SH}(c)}{\partial c}
=
\frac{1}{2c}
\left[
p
-
n_s
E_B
\left\{
\frac{B}{c(1-B)+B}
\right\}
\right].
\label{eq:app_rsh_derivative}
\end{equation}
At \(c=1\), the expectation in \eqref{eq:app_rsh_derivative} becomes
\[
E_B
\left\{
\frac{B}{(1-B)+B}
\right\}
=
E(B).
\]
Using \eqref{eq:app_beta_mean},
\[
E(B)=\frac{p}{n_s}.
\]
Hence
\[
\left.
\frac{\partial \mathcal{R}_{SH}(c)}{\partial c}
\right|_{c=1}
=
0.
\]

We now show that this stationary point is unique. Define
\[
g(c)
=
E_B
\left[
\frac{B}{c(1-B)+B}
\right].
\]
Then
\[
g'(c)
=
-
E_B
\left[
\frac{B(1-B)}
{\{c(1-B)+B\}^2}
\right]
<0,
\]
so \(g(c)\) is strictly decreasing in \(c\). Moreover,
\[
\lim_{c\downarrow 0}g(c)=1,
\qquad
\lim_{c\to\infty}g(c)=0.
\]
Since \(n_s>p\), the term \(p-n_s g(c)\) is negative as \(c\downarrow 0\) and
positive as \(c\to\infty\), and it crosses zero exactly once. Therefore the
first-order condition has a unique solution. Since \(c=1\) satisfies the
first-order condition, it is the unique minimiser of \(\mathcal{R}_{SH}(c)\). Thus, under the Sellentin--Heavens likelihood, the likelihood-loss optimal scalar
covariance correction is
\[
c_{SH}^\star=1.
\]
Unlike the Gaussian plug-in likelihood, the covariance-marginalised likelihood
does not favour an additional global scalar inflation of the sample covariance
matrix.

\end{document}